\documentclass{article}
\usepackage{fullpage}
\newcommand\authorrefmark[1]{$^{#1}$}
\def\address[#1]#2{\\$^{#1}$#2}
\newenvironment{keywords}{\noindent\textbf{Keywords:}}{}
\usepackage{cite,epigraph}
\usepackage{amsmath,amssymb,amsfonts}
\usepackage{algorithmic}
\usepackage{graphicx}
\usepackage{textcomp}
\usepackage{ragged2e} 
\usepackage{colortbl}
\newtheorem{definition}{Definition}[section]
\let\citet=\cite
\usepackage[hyphens]{url}
\usepackage{verbatim}
\usepackage{hyperref}
\usepackage{soul}
\usepackage{versions}\excludeversion{ignore}
\usepackage{comment}

\usepackage{threeparttable}
\usepackage{multirow}
\usepackage{caption}
\begin{document}
\title{\bf Domain-Independent Deception:\\ Definition, Taxonomy and the Linguistic Cues Debate}
\author{{Rakesh M. Verma}\authorrefmark{1} 
\and{Nachum Dershowitz}\authorrefmark{2}\and {Victor Zeng}\authorrefmark{3}  
\and {Xuting Liu}\authorrefmark{4}
\\\address[1]{Department of Computer Science, University of Houston, Houston, TX 77204, USA \\(e-mail: \url{rmverma2@central.uh.edu})}
\address[2]{School of Computer Science, Tel Aviv University, Ramat Aviv, Israel \\(e-mail: \url{nachum@tau.ac.il})}
\address[3]{Department of Computer Science,
University of Houston, Houston, TX 77204, USA \\(e-mail: \url{vzeng@uh.edu})}
\address[4]{Computer Science Department, University of California, Berkeley, CA 94720, USA \\(e-mail: \url{xutingl@berkeley.edu})}
}


\date{}
\maketitle
\begin{abstract}
    Internet-based economies and societies are drowning in deceptive attacks. These attacks take many forms, such as fake news, phishing, and job scams, which we call ``domains of deception.'' Machine-learning  and natural-language-processing  researchers have been attempting to ameliorate this precarious situation by designing domain-specific detectors. Only a few recent works have considered domain-independent deception. We collect these disparate threads of research and investigate domain-independent deception along four  dimensions. First, we provide a new computational definition of deception and formalize it using probability theory. Second, we break down deception into a new taxonomy. Third,  we analyze the debate on linguistic cues for deception and supply guidelines for systematic reviews. Fourth, we provide some evidence and some suggestions for domain-independent deception detection. 
\end{abstract}

\begin{keywords}
Automatic/computational deception, 
Cross domain, Domain independent,  
Email/message scams,
Fake news, 
Meta-analysis,  
Opinion spam, 
Phishing,
Social engineering attacks,
Systematic review,
Text analysis
\end{keywords}

\maketitle

\section{Introduction} \label{sec:introduction}
Trust in Internet-dependent societies and economies is rapidly eroding  due to the proliferation of deceptive attacks, such as fake news, phishing, and disinformation. The situation has deteriorated so much that a significant fraction of the US population believes the 2020 US election was stolen, and a deliberate disinformation campaign contributed to Brexit. 
 
Social-media platforms in particular have come under severe scrutiny regarding how they police content~\cite{gordon19,Verge}. Facebook and Google are partnering with independent fact-checking organizations that typically employ manual fact checkers. Natural-language processing (NLP) and machine learning (ML) researchers have joined the effort to combat this dangerous situation by designing fake news~\cite{shuSW17,zhouZ20}, phishing~\cite{vermaM19,zhouV20,fatiVM22}, and other kinds of domain-specific detectors. 

We think that building single-domain detectors may  be less than optimal. Composing them sequentially requires more time and composing them in parallel requires more hardware. Moreover, building single-domain detectors means we can only react to new forms of deception after they emerge. Hence, we promote designing and building   domain-independent deception detectors. 

Unfortunately, research in this area is currently hampered by the lack of (A) good definitions and taxonomy, (B) high quality datasets, and (C) systematic approaches to domain-independent deception detection. Thus, results are neither generalizable nor reliable, leading to much confusion. 

Our goal is to spur research on \emph{domain-independent} deception by clarifying the key issues and filling  important gaps in the current debate on this topic. Accordingly, we make the following contributions:

\begin{enumerate}
    \item We propose a new computational definition of deception and give a formalization of it based on probability theory\footnote{When we use the term ``deception'' without qualifying adjectives, we mean domain-independent deception. When authors have not been careful about the goals of the deception, we will use the term ``lies'' instead.}  (Section~\ref{sec-defn}).
    \item We give a new taxonomy for deception  that clarifies its explicit and implicit elements  (Section~\ref{sec-defn}). 
    \item We examine the debate on linguistic deception detection, identify works that demonstrate the challenges that must be overcome to develop domain-independent deception detectors, and examine them critically. 
\end{enumerate}

We hope that this article, besides scrutinizing the claims on general linguistic signals for deception, will aid those planning to conduct systematic reviews of the literature. A Google Scholar search with phrase queries of the form ``guidelines for systematic literature reviews in X,'' where X $\in \{$~natural  language processing, NLP,  machine learning$~\}$ 
returned nothing.\footnote{When we dropped ``in X'', we did get some guidelines for reviews of software engineering literature~\cite{kitchenhamC07}, agile software, and the like.}  

The article is organized as follows. In the rest of this section, we give a new computational definition, its formalization and a new, taxonomy of deception.  In Section~\ref{sec-related}, we present the related work on domain-independent deception. In Section~\ref{sec-debate}, we examine the debate on linguistic cues for deception detection. Section~\ref{sec-args} presents our arguments for markers of domain-independent deception. 
The final section presents conclusions and some directions for future work. 

\subsection{Definition} \label{sec-defn}

We first examine a general definition of deception, taken from~\cite{galasinski2000language}, intended to capture a wide variety of deceptive situations and attacks. 
\newline
\begin{definition}[Preliminary]
\textit{Deception}  is an intentional act of manipulation to gain compliance. Thus, it has at least one source, one target, and one  goal. The source is intentionally manipulating the target into  beliefs, or actions, or both, intended to achieving the goal(s). \\
\end{definition}
\noindent
Here, ``compliance'' is with the goals/desires intended by the source. We refer the reader to~\cite{galasinski2000language} for a discussion on several alternate  definitions of deception and their pros and cons. 

Since we are interested in automatic verifiability, we modify this definition of deception and propose one that is computationally feasible. 
As intentions are notoriously hard to establish, we replace it with an exposure of manipulation/goal(s) clause.  
Our revised definition is the following:
\newline
\begin{definition}[Deception]
\textit{Deception} is an act of manipulation designed to gain compliance such that, were the manipulation or the goal(s) of compliance  exposed,  the chances of compliance would decline significantly. 
Thus, it has at least one source, one target, and one goal. 
The source is manipulating the target into  beliefs, or action, or both, intended to achieve the goal(s). \\
\end{definition}

One might require that the goal(s) be harmful to an individual or an organization. However, this would necessitate either a computational definition of harm, or a comprehensive list of potential harms, which could be checked computationally, and is therefore a less desirable alternative. Of course, there remains the task of showing a significant decline in the compliance probability. We argue that this would be usually evident from the exposure of manipulation and/or the goal(s). If the manipulation is non-trivial or the goal(s) malicious, then we expect a significant decline in the probability of success of the deceptive attack upon the exposure of the deceptive method(s) or goal(s). Even if the goals are relatively benign, e.g., satire, parody, etc., the exposure of the goal(s) would reduce the humor or the surprise element or both, which is the compliance in this case, engendered by the deception.

This definition of deception can be formalized using probabilities as follows. In the following definitions, $a$ denotes a deceptive action, $K(a)$ denotes the action $a$ with a clear explanation/explication of the manipulation/deception in $a$ (so in particular $a$ is contained in $K(a)$). We use $T$ to denote the target of action $a$. In probabilistic terms, $a$ is an event. Let $P_T(compl ~|~ a)$ denote the conditional compliance probability of $T$ given $a$ and similarly define $P_T(compl ~|~ K(a))$. Given these notations, we first define reasonable targets of deception, which could be human or automatic detectors. 

By reasonable, we mean to rule out the trivial detectors that use coin tossing or perhaps ``read the tea leaves'' to make a prediction. In other words, a reasonable detector's probability of detection must be dependent on $a$,\footnote{It could also be dependent on the context of the action $a$, but we do not insist on this.}. This can be formalized as the more explicit the deception in $a$ (and correspondingly lesser is the explanation of the manipulation/deception in $a$ needed to be revealed to the detector), the higher is the probability of detection and the lower is the probability of compliance. Therefore, explicit deception (which really is no deception) is when $K(a) = a$.\newline
\begin{definition}
A target $T$ is reasonable provided: (a) $P_T(compl ~|~ a) \not= P_T (compl)$ and (b) $P_T(compl ~|~ a) \geq P_T(compl ~|~ K(a))$ and (c) $P_I(compl ~|~ a) \rightarrow 0$ as $K(a) \rightarrow a$. \\
\end{definition}
This definition means that the probability of compliance is dependent on the action $a$, decreases with explanation of the deception in action $a$, and tends to 0 as the deception in $a$ becomes more explicit, so that no further explanation is required. Now, we can formalize deception. \newline
\begin{definition}[Deception -- Formalized] \label{def:formal}
Given a positive threshold $ 0 < \tau < 1$, we say  an action $a$ is $\tau$-deceptive provided there exists a reasonable target $T$ such that $P_T(compl ~|~ a) - P_T(compl ~|~ K(a)) \geq \tau$, for all actions $a$ that satisfy $K(a) \not= a$.\\
\end{definition}

 In Definition ~\ref{def:formal}, we do {\em not} insist that all reasonable targets satisfy this condition, just one is sufficient. In principle, we could make this formal definition dependent on a given $T$ and compare the threshold required for a given $T$ versus for the threshold for any $T$. This would give us a formalization of the susceptibility of a given target $T$. Another direction would be to eliminate the threshold $\tau$ by just requiring that $P_T(compl | a) > P_T(compl | K(a))$. 
 
There is some work on finding how good humans are at detecting certain kinds of deceptive attacks in~\cite{baki2017scaling,bakiVG20} and for the detection capabilities of automatic detectors one can look at surveys on fake news detection~\cite{sharma2019combating,zhouZ20} and phishing detection~\cite{dasBA20,aassalBDV20} to name a few. 
\subsection{Taxonomy}

Before constructing our own taxonomy based on the new definition of deception, we searched for previous ones. There have been a few attempts at constructing taxonomies for fake news or other forms of deception, e.g., phishing. 
Molina et al.~\cite{molina2019fake} give a taxonomy of {\em fake news}  with four dimensions, viz., (i) message and linguistic, (ii) sources and intentions, (iii) structural, and (iv) network. 
Kapantai et al.~\cite{kapantai2021systematic} make a valiant effort to come up with a unified taxonomy of {\em disinformation}. 
They conducted a systematic search for papers proposing taxonomies for disinformation and synthesized a taxonomy with three dimensions: (i) facticity, (ii) motivation, and (iii) verifiability. 

Yet no one, to our knowledge, has given a comprehensive taxonomy for real-world deception.\footnote{Artificial laboratory situations in which participants are compelled to lie for collecting datasets or some other purpose can also be handled without much modification.} We put forward such a multi-dimensional taxonomy.

To begin our taxonomy, we go back to our definition. From the definition we can see that deception explicitly involves these four  elements, viz., (1) the agents: the source(s) and the target(s), (2) the stratagem(s) for manipulation, (3) the goal(s), and (4) the threat/mechanisms of exposure. Next, there are four concepts that are implicit in the definition: (1) the motivation(s) behind the goal(s), (2) the communication channel(s) or media, (3) the modality of deception, and (4) the manner or timeliness of the exchange. 

We further explain the components below starting with the explicit components.
\begin{enumerate}
\item \textit{Agents}
\begin{enumerate}
\item \textit{Source(s)}. This includes: human (individual or group), bot, etc.,  or mixed, in other words, combinations such as human assisted by a bot.
\item \textit{Target(s)}. This includes: human (individual or group), automatic detector, or both. For example, spam targets automatic detectors, and phishing targets the human but needs to fool automatic detectors also. 
\end{enumerate}
    \item \textit{Stratagem(s)}. This includes: falsification, distortion, taking words out of context, persuasion techniques, and combinations thereof. These are all collectively referred to as implicit or explicit misrepresentations. Persuasion techniques include: authority, scarcity, social proof, reward claims, etc. (see~\cite{bakiV21,bakiV22}).
    \item \textit{Goal(s)}: 
    \begin{enumerate}
        \item \textit{Harmless}: satire, parody, satisfying participation, as in a laboratory experiment where participants may be asked to lie, etc.
        \item \textit{Harmful}: This includes a wide range of objectives, such as stealing money or identity information, malware installation, manipulation of vote, planting fear, sowing confusion, initiating chaos, gaining an unfair edge in a competition (e.g., swaying opinions and preferences on  products), persuading people to take harmful actions, winning competitions/games, etc.
    \end{enumerate}  
    
    \item \textit{Exposure}
    \begin{enumerate}
    \item \textit{Facticity}. Can we establish whether it is factual or not? For example, currently we are unable to establish the truth or falsity of utterances such as, ``There are multiple universes in existence right now.'' 
    \item \textit{Verifiability}. Assuming facticity, how easy or difficult it is to verify whether it is legitimate or deceptive? Here, we are interested in machine or automatic verification. If a simple machine-learning algorithm can detect it with high (some threshold, say over 95\%) recall and precision, we will deem it easy. 
    \end{enumerate}
\end{enumerate}

\begin{figure*}{}
\centering
\includegraphics[width=0.9\textwidth]{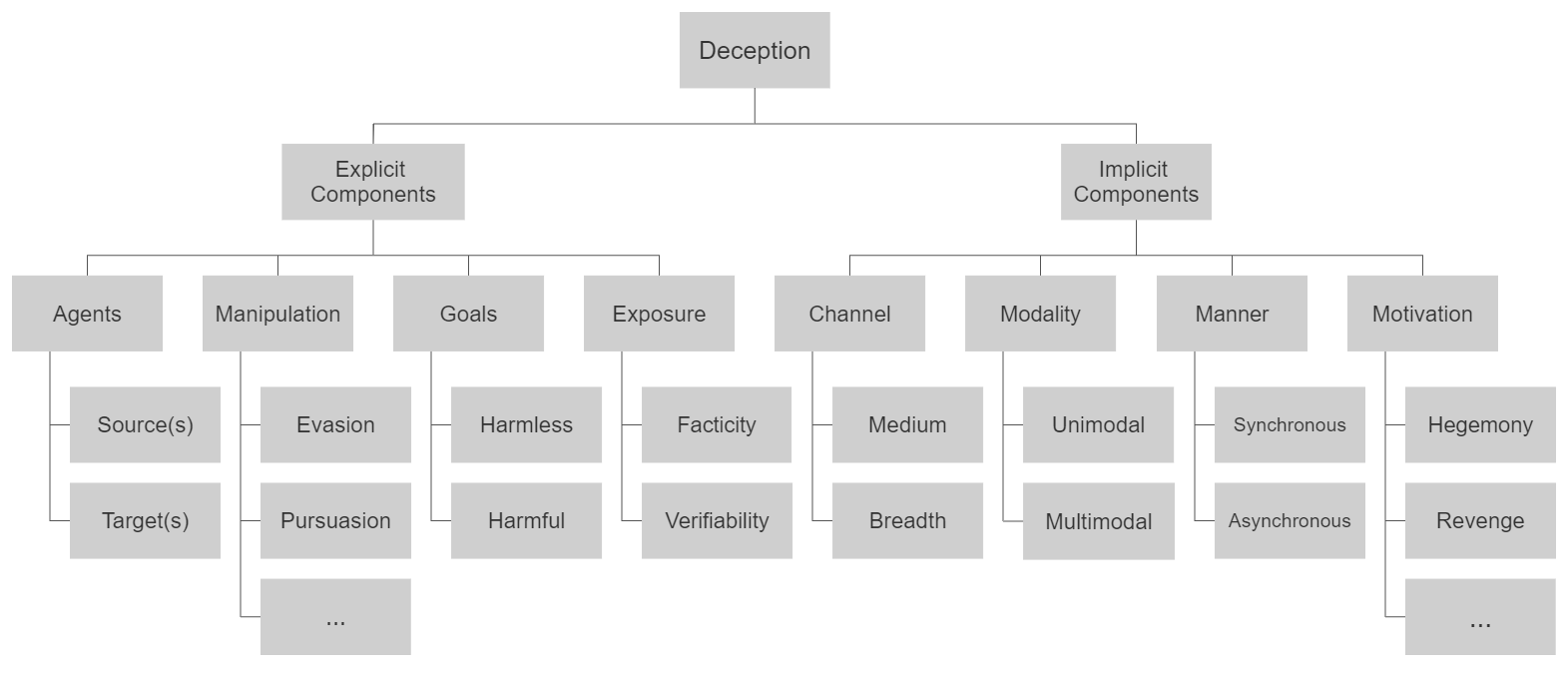}
\caption{The Deception Taxonomy} \label{fig:taxonomy}
\end{figure*}

Next, we examine the implicit components of deception. 
\begin{enumerate}
\item \textit{Motivation}. This is the rationale for the goal(s). The agents involved and their characteristics reveal the underlying motivations, which could be political hegemony (nation states), religious domination, revenge (disgruntled employee), ideological gains, etc. 
    \item \textit{Channel}. This dimension includes two aspects:
    \begin{enumerate}
        \item Breadth: Whether the targets are a few specific individuals or broad classes of people.
        \item Medium: How is the deceptive capsule conveyed to the target(s).
    \end{enumerate} 
    \item \textit{Modality}. This dimension refers to the presentation of the deceptive content. It includes:
    \begin{enumerate}
        \item Unimodal: This includes only  one type of modality such as (a) Gestural: body language is used to deceive, (b) Audio  (a.k.a.\@ verbal), (c) Textual (e.g., SMS/email), and (d) Visual  (e.g., images or videos).
        \item Multimodal: Combinations of different modalities.
    \end{enumerate}  For example, audio-visual means it has both speech and visual components, but lacks face-to-face communication in which gestures could be used to facilitate deception.
    \item \textit{Manner/Timeliness}. 
    \begin{enumerate}
        \item Interactive/Synchronous: A real-time interview or debate is an interactive scenario.
        \item Noninteractive/Asynchronous:  An Amazon Mechanical Turker  typing a deceptive opinion or essay is a non-interactive one. 
    \end{enumerate} 
    \end{enumerate}
Our taxonomy is shown in Figure~\ref{fig:taxonomy}.

To the best of our knowledge, the following dimensions are new in the above taxonomy: target, stratagem, goal, dissemination, and timeliness. We add these to give a comprehensive view of deception, to aid in domain-independent deception detection, and to clarify and classify deception in all its different manifestations.
Such a comprehensive taxonomy will provide a solid foundation on which to build automatic and semi-automatic detection methods and training programs for the target(s) of deception.

We use the term ``domain'' to refer to  the \emph{goal} of the deception. Therefore, by ``domain independence'' we mean that the goals of deception can be quite different. 

\section{Related Work on Domain-Independent Deception} \label{sec-related}
The related work on deception detection can be categorized into: datasets, detection, and literature reviews. Of the last category, we focus on reviews of the linguistic deception detection literature here. The DBLP query, ``domain decepti'' (since authors can use ``deception'' or ``deceptive''), on 9 November 2021, gave 16 matches of which nine were deemed relevant. 

\noindent {\bf Remark.} Unfortunately, previous researchers have generally left the term ``domain'' undefined. 
In~\cite{glenski2021towards}, researchers refer to different social networks as domains, e.g., Twitter versus Reddit. 
Hence, terms such as ``cross-domain deception'' in previous work could mean that the topics of essays or reviews are varied but the goal(s) could stay pretty much the same.

\subsection{Datasets}\label{sec-data}
Several datasets have been collected for studying lies. However, researchers have not carefully delineated the scope by considering the goals of the deception.  For example, Zhou et al.~\cite{zhou} paired students and asked one student in each pair to deceive the other using messages. There are multiple datasets for fake news detection, opinion spam (a.k.a.\@ fake reviews) detection, and for phishing~\cite{ELAetAl:IWSPA-AP2018}. Next, we discuss  datasets, where the term ``domain'' is used in the topic sense.\footnote{Note that the topics can vary in a heterogeneous application, e.g., fake news detection, since some items could be on sport and some on politics or religion. Moreover, the goals could be different too. Hence, we will not use the term ``domain'' to refer to applications such as fake news.} 

In~\cite{peres-rosas14}, researchers collected demographic data and  14 short essays (7 truthful and 7 false) on  open-ended topics by 512  Amazon Mechanical Turk workers. We refer to this as the \textit{Open-Domain} dataset. They tried to predict demographic information and facticity. In~\cite{perez2014cross}, researchers collected short essays on three topics: abortion, best friend, and  death penalty by people from four different cultural backgrounds. The definition of domains in these two papers, viz., topics, is finer-grained than ours, wherein the goal(s) of deception are varied, not just topical content.  

In~\cite{Hernandez-CastanedaCG17}, researchers analyzed three datasets: a two class, balanced-ratio 236 Amazon reviews dataset, a hotel opinion spam dataset consisting of 400 fabricated opinions from Mechanical Turkers plus 400 reviews from TripAdvisor (likely to be truthful), and 200 essays from~\cite{perez2014cross}. In~\cite{XarhoulacosASG21}, researchers studied a masking technique on two datasets: a hotel, restaurant, and doctor opinion spam dataset and the dataset from~\cite{perez2014cross}. The idea is to mask the content that is not relevant in deception detection. In~\cite{cagnina2017detecting}, in-domain experiments were done with a positive and negative hotel opinion spam dataset, and cross-domain experiments were conducted with the hotel, restaurant, and doctor opinion spam dataset. In~\cite{capuozzo}, truthful and deceptive opinions on five topics are collected in multiple languages. 
In these papers also we see that the topics of deception are varied rather than the goal(s).

To our knowledge, the following works have developed  domain-independent deception datasets in our sense, wherein the goal(s) of deception can be quite different:~\cite{Rill-GarciaPRE18,shahriar2021domain,vogler2020using,XarhoulacosASG21, yeh2021lying}. 

In~\cite{Rill-GarciaPRE18}, researchers used two datasets: the American English subset of~\cite{perez2014cross} consisting of a balanced-ratio 600 essays  and transcriptions of 121 trial videos (60 truthful and 61 deceptive), which we call Real-Life\_Trial below.  
In~\cite{vogler2020using}, researchers used three datasets: positive and negative hotel reviews, essays on emotionally-charged topics, and personal interview questions. 
In~\cite{XarhoulacosASG21}, multiple fake news datasets, a Covid-19 dataset, and some micro-blogging datasets are collected  and analyzed. 
In~\cite{shahriar2021domain}, researchers collected fake news, Twitter rumor and spam datasets.\footnote{Spam is essentially advertising and deception is employed to fool automatic detectors rather than the human recipient of the spam. We focus more on human targets in this work.} They applied their models trained on these datasets to a new Covid-19 dataset.  

In~\cite{yeh2021lying}, researchers collected seven datasets (Diplomacy, Mafiascum, Open-Domain, LIAR, Box of Lies, MU3D, and Real-Life\_Trial) and analyzed them using LIWC categories. In~\cite{yeh2021lying},  researchers do not claim domain independence or cross-domain analysis. However, their datasets do involve different goals, e.g., LIAR includes political lies with the goal of winning elections, whereas the lies in Real-Life\_Trial have other goals, and Diplomacy/Mafiascum are about winning online games. 

It is clear from the above discussion that we still lack large, comprehensive datasets for deception that have a wide variety of deceptive goals.

\subsection{Detection}

Deception detection in general is a useful and challenging open problem. There have been many attempts at specific applications such as phishing and fake news. On phishing alone (query: phish), there are more than 1,100 DBLP results, including at least 10 surveys and reviews. Similarly, there are 419 papers on scams, 81 on opinion spam, 74 on fake reviews, and 536 on fake news.

The only works on domain-independent deception detection are already covered in the Datasets section. However, there are some reviews, surveys,  and meta-analysis of linguistic cues for deception detection. 

\subsection{Reviews on Linguistic Deception Detection}

Recently,   Gröndahl  and   Asokan~\citet{grondahlA19} conducted a survey of the literature on deception. They defined implicit and explicit deception,\footnote{Explicit deception is when the deceiver explicitly mentions the false proposition  in the deceptive communication.} focused on automatic deception detection using input texts, and then proceeded to review 17  papers on \textit{linguistic} deception detection techniques. These papers covered two forms of deception: (a) dyadic pairs in the laboratory, where one person sends a short essay or message to another (some truthful and some lies), and  (b) fake reviews (a.k.a.\@ opinion spam).  

Based on their analysis of the literature on laboratory deception experiments and the literature on opinion spam, they concluded that \textit{there is no linguistic or stylistic trace that works for deception in general}. 

Similarly, the authors of~\citet{vogler2020using} assert that extensive psychology research~\cite{vrij08, fitzpatrickBF15} shows that ``a generalized linguistic cue to deception is unlikely to exist.''

In subsequent sections, we collectively refer to the deception survey of~\citet{grondahlA19} and the triad of papers~\cite{fitzpatrickBF15,vogler2020using,vrij08} as 
the \textit{Critiques}.
We argue that, at best, their analyses and conclusion may be a bit too hasty. 
We elaborate on several aspects that need further and deeper investigation/analysis with specific examples from the reviewed literature. 
Although we focus on those specific critiques here, many of the issues we raise are more generally applicable to any systematic review of scientific literature. 

\section{The Debate on Linguistic Cues for Deception Detection} \label{sec-debate}
We begin with some general guidelines for systematic reviews and then focus on the debate on linguistic cues for domain-independent deception detection. 
\subsection{Guidelines for Systematic Reviews}
One may observe a recent explosion in systematic reviews on all kinds of problems in natural language processing  and machine learning. 
A good systematic review can benefit both academics and professionals by organizing the literature, highlighting key results and identifying gaps in our understanding.
However, there is a dearth of good guidelines and procedures for such systematic reviews in NLP and computer science in general. 


According to~\citet{staplesN07}, ``A systematic review is a defined and methodical way of
identifying, assessing, and analyzing published primary
studies in order to investigate a specific research question.''
Such a review can reveal the structure and
patterns of existing research, and  identify gaps for future research~\cite{kitchenhamC07,staplesN07}. 
Furthermore,
per~\cite{staplesN07}, systematic reviews are 
formally planned and methodically executed. 

\subsubsection*{Evaluating Reviews}

A good systematic review is independently replicable, and thus has additional scientific value over that of a literature survey.  In collecting, evaluating, and documenting all available evidence on a specific research question, a systematic review may provide a greater level of validity in its findings than might be possible in any individual study reviewed. 
However, systematic reviews require much more effort than ordinary literature surveys.

The following features differentiate a systematic review
from a conventional one (Kitchenham, 2004):
\begin{itemize}
\item A pre-defined and documented protocol specifying the research question and procedures to be used in performing the review.
\item A defined and documented search strategy designed to find \textit{as much of the relevant literature as possible}.
\item Explicitly pre-defined criteria for determining whether to include or exclude a candidate study. 
\item Description of quality assessment mechanisms to evaluate each study.
\item Description of review and cross-checking processes  involving multiple independent researchers, to control researcher bias.
\end{itemize}
\subsection{The Debate on Linguistic Cues}
The deception survey of~\citet{grondahlA19} has a few of these features: they specify the research questions and hypotheses and involve two researchers (presumably mentor and mentee). 
However, no review protocol is presented, no search strategy is defined, no inclusion/exclusion criteria are explicated, and no quality assessment mechanism is specified. 
Thus, it is probably better to view their paper as a conventional literature review. 
However, as it is published in an influential journal, it is likely to leave a lasting impression on deception researchers. 
Hence, we believe it is worth the time and effort to examine its strengths and weaknesses more closely. 

The statements of~\cite{vrij08} and~\cite{fitzpatrickBF15} stem from a meta-analysis conducted by~\cite{depauloLM03}. 
On the positive side, a meta-analysis can be very useful and adds statistical analysis methods to a systematic review. 
However, this meta-analysis is now quite dated and suffers from confirmation bias among other issues as we outline below. 

Next, we consider the issues and challenges that can arise with systematic reviews in general and then those that are specific to the papers under consideration. 

\subsection{Issues and Challenges} 
We enumerate several challenges with  reviews and surveys, whether they are systematic or conventional, emphasizing those that are common to the Critiques. 

\subsubsection{Publication Bias}
Not having a clear, explicit search strategy for literature or clearly defined inclusion and exclusion criteria can lead to a study that displays biases regarding the publications that are covered. 
The deception survey~\cite{grondahlA19} suffers from this issue. Although their goal was to survey automatic linguistic deception detection literature, they  missed not just many relevant papers, including~\cite{duranCHMM09,fullerBW09,levitanMH18,schellemanM10,zheng18}, but also the meta-analysis of~\cite{hauch16}.\footnote{Since the~\cite{grondahlA19} paper was revised in 2019, we used 2018 as the cutoff for listing missing literature.} 

This meta-analysis examined 79 cues from 44 different studies on automatic linguistic deception detection. They state: ``The meta-analyses demonstrated that, relative to truth-tellers, liars experienced greater cognitive load, expressed more negative emotions, distanced themselves more from events, expressed fewer sensory–perceptual words, and referred less often to cognitive processes. However, liars were not more uncertain than truth-tellers. These effects were moderated by event type, involvement, emotional valence, intensity of interaction, motivation, and other moderators. Although the overall effect size was small, theory driven predictions for certain cues received support.'' 

However, there is another, more serious issue. Reviewing only papers that are published means ignoring papers that remain unpublished for one reason or another. For example, positive studies are more likely to be published than negative studies, papers in English have a higher chance of being published as also papers authored/coauthored by researchers who are already highly reputed, etc. Moreover, longer works such as theses and dissertations are also missed in the emphasis on published literature. 

To study if this bias exists in the Critiques, we did a systematic search of the ProQuest Global Database~\cite{proquest}. 
We identified 117 dissertations and theses with the keywords ``deception'' \textit{and}  ``detection'' in the title. 
Three of these also have  the word ``linguistic'' in the title and all three are relevant~\cite{fuller08,hauch16,humpherys10}.\footnote{There are 34 with ``deception'' and ``detection'' in the title and ``linguistic'' anywhere in the document.} 
Replacing linguistic with ``natural language processing'' (or ``textual'') and keeping ``deception'' yielded two more relevant  dissertations~\cite{Cooper08,picornell13}. 
Finally, ``verbal'' with ``deception'' yielded four more relevant  results, viz.,~\cite{akehurst97,perez10,riggio81,vartapetiance15}, out of nine total. 
The  last three searches looked for both keywords in the title only. 

None of the above dissertations are cited in the Critiques, although a paper by the author of~\cite{humpherys10} does appear in~\cite{fitzpatrickBF15} and the author of~\cite{akehurst97} appears as a coauthor on some papers cited in~\cite{vrij08}. 
The meta-analyses of~\cite{depauloLM03}, with more than 80 studies  covered, and of~\cite{hauch16}, 44 studies  covered, are much more systematic and comprehensive. 
For example, the queries used are listed in~\cite{hauch16} and the inclusion and exclusion criteria are spelled out. Publication bias is mentioned in the meta-analysis of~\cite{hauch16} and an effort is made to select independent studies. 
However, the meta-analysis of~\cite{depauloLM03} also did not check for publication bias. 

\begin{table}{} 
\centering 
\caption{The nine most prolific authors in the textual deception detection literature studied in~\cite{grondahlA19}.} \label{tab-multiple}
\begin{tabular}{|c|c|} 
    \hline
    \bf Author & \bf Papers \\ \hline
    Judee K. Burgoon & 4 \\
    Jeffrey T. Hancock & 4 \\
    Jay F. Nunamaker Jr. & 3 \\
    Lina Zhou & 3 \\
    Doug P. Twitchell & 3 \\
    Myle Ott & 3\\
    Claire Cardie & 3\\
    Yejin Choi & 2 \\
    Tiantian Qin & 2 \\
    \hline
    \end{tabular}
\end{table}

\begin{figure}{}
\centering
\includegraphics[width=0.7\columnwidth]{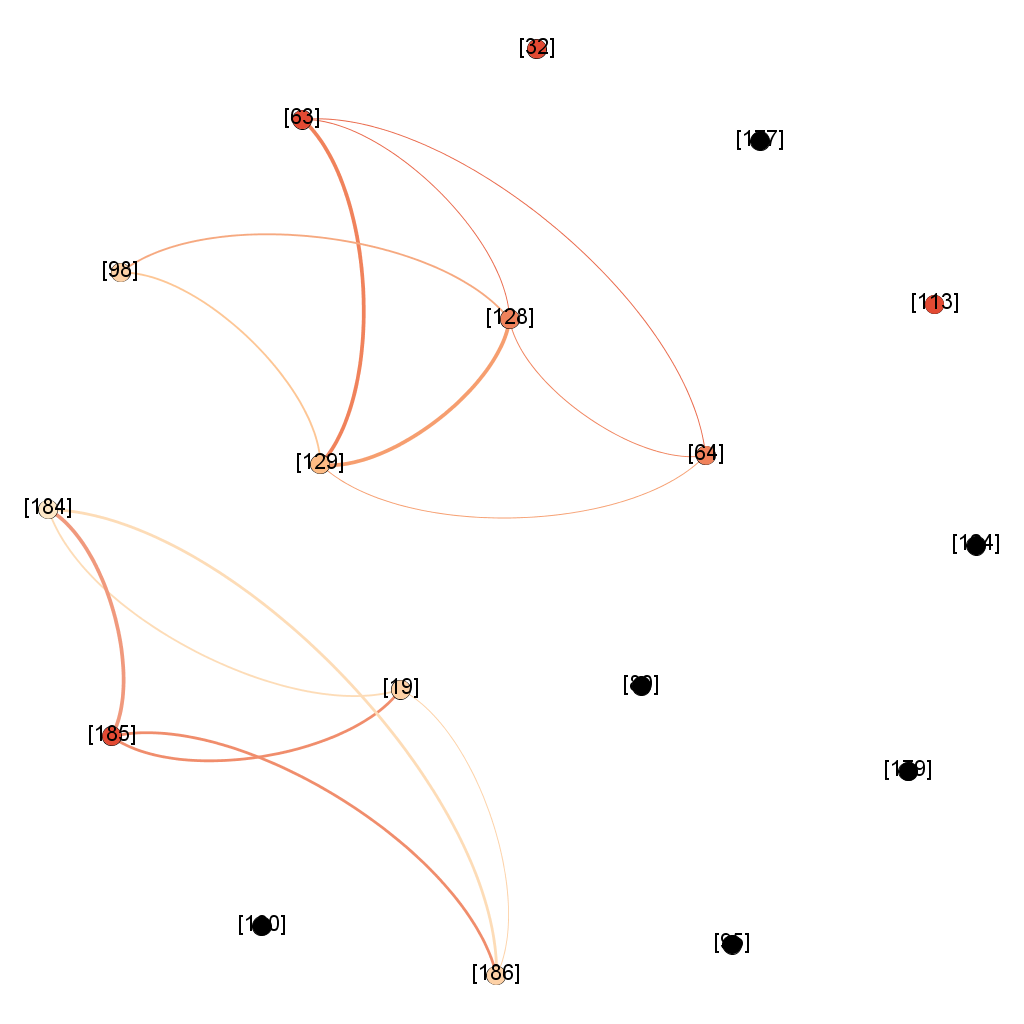}
\caption{Graph showing the 17 papers as vertices. There is an edge between two papers (vertices) provided they have a common author. The thickness of the edge and color reflect the number of common authors. Several cliques ($K_3$ and $K_4$) are visible.} \label{fig-graph}
\end{figure}

There is another kind of publication bias, when a review focuses too much on papers from a clique of interconnected researchers, or papers that analyze the same dataset multiple times. 
We  observe this kind of bias in the deception survey~\cite{grondahlA19}. 
To see this, we listed all the authors of the 17 papers cited in Section 2 (``Deception Detection Via Text Analysis'') of their paper. 
There are 56  authors in total, but only 38 unique names. 
We report all authors with more than one publication in Table~\ref{tab-multiple}. Next, we analyze papers with common authors in the graph of Fig.~\ref{fig-graph}. It shows several cliques, two of them as large as a $K_4$ (complete graph on 4 vertices) and one $K_3$ that is {not} a subgraph of the $K_4$'s. 

\subsubsection{Confirmation Bias}
Next, we discuss perhaps the most serious issue with the Critiques, confirmation bias.

None of the papers examined in the meta-analysis conducted by~\cite{depauloLM03} and the deception survey by~\cite{grondahlA19} built a general dataset for different deception goals (e.g., as in phishing, fake news, \textit{and} crime reports). 
If researchers study a particular form of deception and build a dataset to study it, the chance that they would stumble upon general linguistic cues for deception is likely to be small, since that was not even their objective anyway! 
Hence, a review of these papers is also unlikely to find any general linguistic cues for deception. 

\subsubsection{Quality of Studies and Datasets}
Another issue is the quality of the studies and the datasets collected. 
Quality of studies includes many different factors such as: (i) the design of the experiments, (ii) the sizes and the heterogeneity of the populations, (iii)   whether the statistical tests used are appropriate for the datasets analyzed, whether tests of statistical significance were applied and correctly reported so that effect sizes can be obtained, (iv) whether something like the Bonferroni-Holm~\cite{aickinG96}  correction was used for the multiple comparisons issue, and (v) their replicability. 

Another issue that must be considered is whether the participants, typically undergraduates, in laboratory experiments are as motivated as real-world attackers in carrying out the deception~\cite{hartwigB14}. 

\subsubsection{Datedness}
The meta-analysis of~\cite{depauloLM03} was conducted in 2003. 
The meta-analysis of~\cite{hauch16} is more recent, but still only covers papers up to February 2012. 
The latest review of meta-analyses~\cite{sternglanzMM19} on deception detection lists more than 50 meta-analyses.
Of course, not all are relevant to linguistic deception detection, but this points to the large volume of work in the field and is indirect evidence for the inadequacy of the literature cited in the  Critiques. 

 \subsubsection{New Developments in NLP}
 Computer science, machine learning, and NLP have come a long way since 2012. 

Recent breakthroughs such as attention, transformers, and pre-trained language models like BERT, have revolutionized NLP. Even if the previous critiques were valid, their conclusions should be reexamined in light of these new advances. We refer the reader to~\cite{Zeng_codaspy} for efforts in this direction.

\section{Arguments for Domain-Independent Deception Markers} \label{sec-args}
Next, we examine the positive case in favor of existence of general linguistic cues of deception. In contrast to the assertion in the Critiques, there are several arguments in favor of general linguistic markers for deception. 

\subsection{Prior Analyses}

First, the meta-analyses of~\citet{depauloLM03} and~\citet{hauch16} did find markers of deception in the studies they examined.
Although the effect sizes were low to moderate, bear in mind that they conducted these meta-analyses on papers that studied specific forms or situations of deception and did not build any general domain-independent datasets. 

Second, the following papers all point to evidence for cross-domain deception detection:~\cite{Rill-GarciaPRE18,shahriar2021domain,vogler2020using,XarhoulacosASG21,yeh2021lying}. As mentioned above, these researchers have created domain-independent datasets and developed features and techniques for deception detection across domains.

There is much more that could be done. For example, broad studies of deception should also include deceptive attacks such as phishing and social engineering attacks.  

\begin{table}
    \centering 
    \caption{Comparing the latest surveys, reviews and meta-analysis on automatic deception detection. {QL/DB}: Whether the queries/databases searched are listed. {Period}: The time period of searches listed in the paper. {Papers}: Number of papers  surveyed. {Ling?}: Is there support for linguistic features?}
    \label{tab:the_table}
    \begin{tabular}{|c|c|c|c|c|c|}
        \hline
      \bf  Reference &  \bf QL  &  \bf DB & \bf Period & \bf Papers &  \bf Ling? \\\hline
         GA19~\cite{grondahlA19} & No & No & -- & 18 & No \\ \hline
        H16~\cite{hauch16} & Yes & Yes & 2011--12 & 44 & Yes \\ \hline
  E19~\cite{elhadad2019fake} & Yes & Yes & 2017--19 & 47 & Partial \\ \hline
    \end{tabular}
\end{table}

The meta-analysis of~\cite{hauch16} searched four databases: PsycInfo, Social Science Citation Index, Dissertation Abstracts, and Google Scholar for articles between 1945 and February 2012 with ``all permutations and combination of one keyword from three different clusters: (i) verb, language and linguistic; (ii) computer, artificial, software and automatic; (iii) lie, deceit, decept*.''

The systematic review of~\cite{elhadad2019fake} searched Google Scholar for articles between 2017--2019 using 10 queries listed in their paper. Their queries  are a \textit{proper} subset of the Boolean query 
\begin{quote}
(fake $\vee$ false) news (identify $\vee$ detect) on (social media $\vee$ twitter)
\end{quote}
which we repeated on Scholar on 11 November 2021, with a claim of 1,020,000 results.\footnote{Google counts are loose upper bounds of actual matches.} Their queries produced a total of 157 potentially relevant results. We summarize the pertinent characteristics of the three latest reviews/surveys/meta-analysis of deception~\cite{elhadad2019fake,hauch16,grondahlA19} in Table~\ref{tab:the_table}. 

\subsection{Our Analysis}

Since the meta-analysis of~\cite{hauch16} 
ended in February 2012, we searched the following databases: Google Scholar, PsycInfo and Dissertations, and Abstracts Global, for the period 2013--2021, with the query
\begin{quote}
(verbal $\vee$ language $\vee$ linguistic $\vee$ text $\vee$ lexical) $\wedge$ (computer $\vee$ artificial $\vee$ software $\vee$ automatic $\vee$ autonomous $\vee$ automated $\vee$ identify $\vee$ computational $\vee$ machine $\vee$ detect $\vee$~ tool) $\wedge$ (lie $\vee$ false $\vee$ fake $\vee$ deceit $\vee$ deception $\vee$ deceptive)
\end{quote}
We formed this query by appropriately combining the queries from~\cite{hauch16,elhadad2019fake}, adding keywords after scanning the initial results, and adding relevant synonyms from querying WordNet 3.1 with ``deceit'', ``identify'', and ``lexical''. 
Adding ``recognition'' to the  middle clause reduced the set of results by more than 100K, a flaw of Google Search. 
We tried other synonyms such as ``discover'', ``recognize'', and ``recognizing'' for ``identify'' and ``fraud'' for ``deceit'' in separate queries, but their results on Scholar seemed irrelevant. 

Scholar claimed over 1,150,000 results but only displayed the top 1,000 in relevance. A scan through this list identified 880 as potentially relevant matches. PsycInfo gave us 456 matches and the Dissertations database yielded 134 matches. 

A new query was tried on Scholar without limiting the time period: 
\begin{quote}
(verbal $\vee$ language $\vee$ linguistic $\vee$ text $\vee$ lexical) $\wedge$ (computer $\vee$ artificial $\vee$ software $\vee$ automatic $\vee$ autonomous $\vee$ automated $\vee$ identify $\vee$ computational $\vee$ machine $\vee$ detect $\vee$~ tool $\vee$ recognize $\vee$ recognition $\vee$ recognizing) $\wedge$ (rumor $\vee$ hoax $\vee$ misinformation $\vee$ disinformation)
\end{quote}
Scholar claimed 350,000 results and a scan of the top 1000 gave us 186 potentially relevant matches. 

The results from Scholar were searched for feature selection and feature ranking papers. The 175 resulting papers included surveys, dataset, and research papers. All surveys in the last five years were analyzed for insights on features. More than one survey mentioned $n$-grams of part of speech (POS) tags and semantic features~\cite{guo2020future,sharma2019combating,zhouZ20} as examples of generalizable features. However, this analysis also revealed the lack of feature rankings for large, diverse and general datasets of deception.

\section{Conclusions and Future Work} \label{sec-concl}
We have provided new definitions for deception based on explanations and probability theory. We gave a new taxonomy for deception that clarifies the explicit and implicit elements of deception.  We have given sound desiderata for systematic review and meta-analysis, which we hope will help researchers conduct high quality analyses of the literature and devise new domain-independent deception detection techniques. 

We have argued against hasty conclusions regarding linguistic cues for deception detection and especially their generalizability. 
The Critiques contained in~\cite{fitzpatrickBF15,vogler2020using,vrij08} may present a valid point, namely that linguistic cues might not generalize across the broad class of attacks. However, more sweeping statements should be made with caution, as they discourage future domain-independent deception research and there is some evidence to suggest they are likely to be at least partly off the mark~\cite{Zeng_codaspy}.

With all the new developments in machine learning and NLP, we believe that research on linguistic deception detection is poised to take off and could result in significant advances. We believe that there is still much scope for work on linguistic cues for deception, since our analysis has revealed critical gaps in the literature. We hope that our effort will clarify the terminology, scope of deception and give further impetus to domain-independent deception detection.

\section*{Acknowledgments}  Verma's research was partially supported by NSF grants 1433817, 1950297 and 2210198, ARO grant W911NF-20-1-0254, and ONR award N00014-19-S-F009.  He is the founder of Everest Cyber Security and Analytics, Inc. Zeng's research was supported by ONR award N00014-19-S-F009 and Liu's research was supported by NSF award 1950297.

\bibliographystyle{plain}
\bibliography{eacl,ref}

\end{document}